\documentclass[aip,graphicx]{revtex4-1}
\usepackage{amsmath}    
\usepackage{graphicx}   
\usepackage[hang]{subfigure}
\newcommand{\jun}{junction }
\newcommand{\juns}{junctions }
\newcommand{\Jos}{Josephson }
\newcommand{\elli}{elliptic }
\newcommand{\ann}{annular }

\begin{document}
\title[R.Monaco \textit{et al.}]{Elliptic Annular \Jos Tunnel Junctions in an external magnetic field: The dynamics}
\author{Roberto Monaco$^{\dag}$}
\address{CNR-ISASI, Institute of Applied Sciences and Intelligent Systems ''E. Caianello'', Comprensorio Olivetti, 80078 Pozzuoli, Italy}
\email{ r.monaco@isasi.cnr.it}
\author{Jesper Mygind}
\address{DTU Physics, B309, Technical University of Denmark, DK-2800 Lyngby, Denmark}
\email{myg@fysik.dtu.dk}

\date{\today}

\begin{abstract}
We analyze the dynamics of a magnetic flux quantum (current vortex) trapped in a current-biased long planar elliptic annular Josephson tunnel junction. The system is modeled by a perturbed sine-Gordon equation that determines the spatial and temporal behavior of the phase difference  across the tunnel barrier separating the two superconducting electrodes. In the absence of an external magnetic field the fluxon dynamics in an elliptic annulus does not differ from that of a circular annulus where the stationary fluxon speed merely is determined by the system losses. The interaction between the vortex magnetic moment and a spatially homogeneous in-plane magnetic field gives rise to a tunable periodic non-sinusoidal potential which is strongly dependent on the annulus aspect ratio. We study the escape of the vortex from a well in the tilted potential when the bias current exceeds the depinning current. The smallest depinning current as well as the lowest sensitivity of the annulus to the external field is achieved when the eccentricity is equal to $-1$. The presented extensive numerical results are in good agreement with the findings of the perturbative approach. We also probe the rectifying properties of an asymmetric potential implemented with an egg-shaped annulus formed by two semi-elliptic arcs.

\end{abstract}
\maketitle
\tableofcontents
\section{Introduction}

\noindent A planar \Jos tunnel \jun (JTJ) is certainly the best solid-state device for the study of non-linear phenomena and, in particular, for the investigation of soliton dynamics. In this context, a soliton is a current vortex, also called a \Jos vortex or a fluxon carrying one magnetic flux quantum. A fundamental model describing the fluxon dynamics is based on the perturbed sine-Gordon equation with additional terms which account for dissipation and driving fields. It has been recognized a long time ago \cite{scott} that the fluxon(s) motion is smoother in ring-shaped \juns since the collision of the fluxon with the boundaries are absent. Another unique property of not simply-connected \juns is due to the fluxoid quantization in the superconducting loop formed by either the top or the bottom electrodes of the tunnel junction. One or more fluxons may be trapped in the \jun during the normal-superconducting transition. Once trapped the fluxons can never disappear and only fluxon-antifluxon pairs can be nucleated. Later on, many researchers found the circular geometry ideal for experimental tests of the perturbation models developed to take into account the dissipative effects in the sine-Gordon analysis \cite{davidson, dueholm,hue}. Circular annular JTJs, consisting of two superconducting rings coupled by a thin dielectric layer were also recognized to be the ideal device to investigate both the statics and the dynamics of sine-Gordon solitons in the presence of a periodic potential \cite{gronbech,ustinov,PRB98,wallraf3}. A spatially periodic potential for the fluxon can be easily implemented by a magnetic field applied in the \jun plane. However, for JTJs having a not simply-connected topology the most general and at the same time regular geometry is provided by an \elli annulus. At variance with a circle that has infinitely many axes of symmetry, an ellipse has two axes of symmetry.  Elliptic \ann \Jos tunnel junctions (EAJTJs) serve as a handy tool for the realization of complex periodic potentials, including those lacking spatial reflection symmetry, known as ratchet potentials \cite{magnasco}. An additional motivation to study EAJTJs is to cast in one unique class the many apparently different JTJ configurations, including the linear geometry commonly studied in the context of JTJs. Furthermore, by constructing the junction from two halves of different ellipses it is possible to form an asymmetric oval or egg-shaped annulus with just one axis of symmetry. 

\noindent In this article we focus on the dynamic properties of the phase in long EAJTJs; their static properties in the presence of an externally applied field were the subject of a recent work \cite{SUST15a}. The paper is organized as follows. In the next subsection we state the problem by describing what an EAJTJ is and introduce the mathematical notations and identities used in the paper. In Section II we will describe the sine-Gordon modeling for curved long JTJs and examine the partial differential equation appropriate to the specific case of an elliptic annulus in a uniform in-plane magnetic field. In Section III we study the dynamic properties of a single fluxon trapped in a long EAJTJ, particularly when a magnetic field is present; we discuss the results of the numerical simulations and outline the effects of the ellipse eccentricity. Later on, by combining two semi-elliptic \juns with different aspect ratios, we suggest a simple geometrical configuration that implements a periodic ratchet potential. The conclusions are drawn in Section V.

\subsection{Elliptic annular junctions}

\begin{figure}[tb]
\centering
\includegraphics[width=7cm]{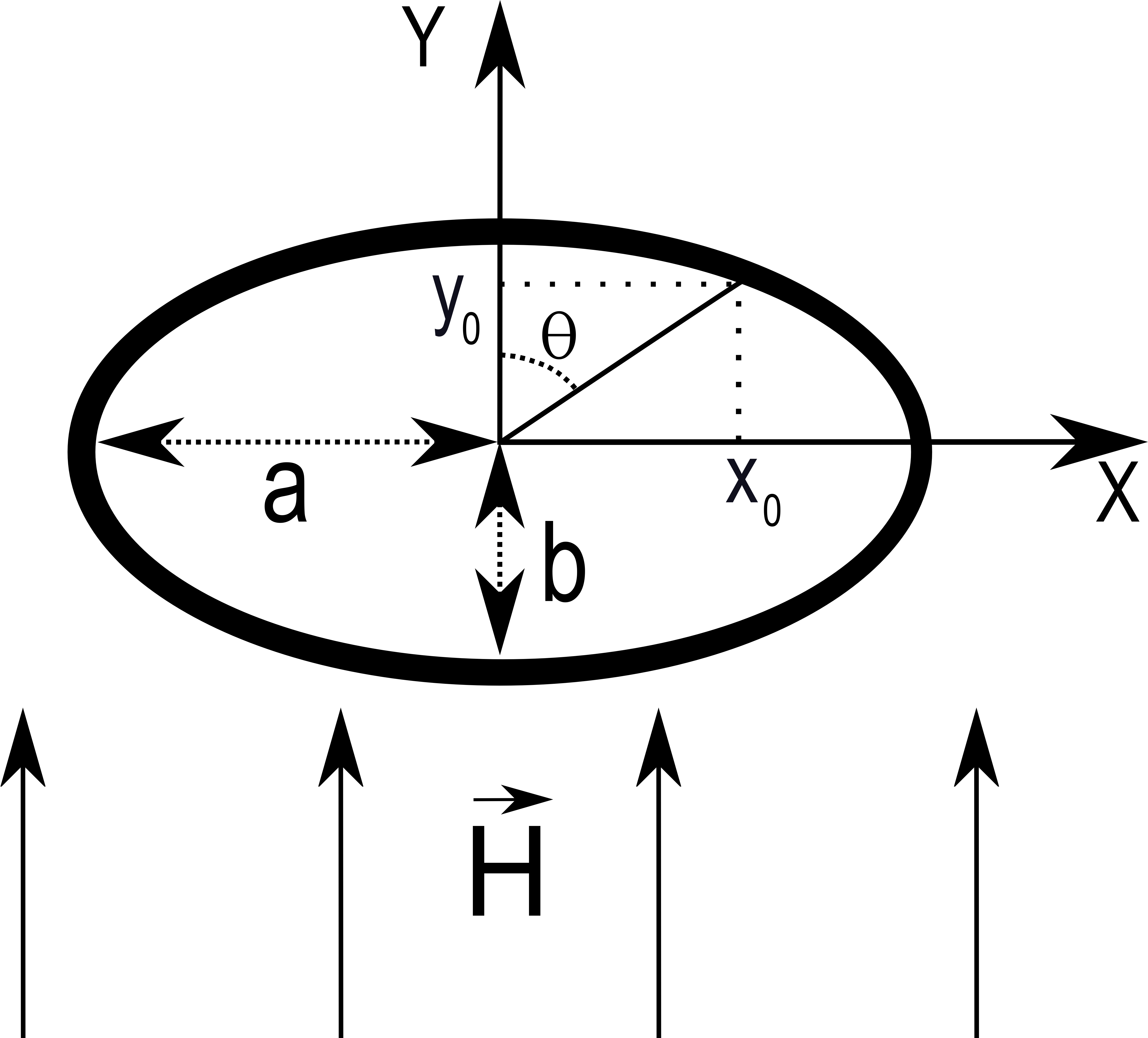}
\caption{Drawing illustrating the tunneling area of an elliptic annular \Jos tunnel \jun in a spatially homogeneous in-plane magnetic field ${\bf H}\equiv \left(0, H \right)$. The annulus width is constant and much smaller than the ellipse semi-axes $a$ and $b$.}
\label{ellipse}
\end{figure}

Let us consider a planar tunnel junction shaped as an elliptic annulus that, as shown in Figure~\ref{ellipse}, lays in the plan identified by the $X$-$Y$ Cartesian coordinate system whose origin coincides with the annulus center of symmetry and whose axes are directed along the principal semi-axes $a$ and $b$. We define the axes ratio $\rho\equiv b/a$ and the eccentricity $e^2 \equiv 1-\rho^2$. We will assume the annulus width to be constant and much smaller than the ellipse semi-axes. The internal and external boundaries of such an annulus can be implemented by drawing two closed curves {\it parallel} to a master ellipse, with constant but opposite offsets; strictly speaking, such curves are not ellipses, but more complex curves \cite{http}. The master ellipse is described by the parametric equations $x=a\sin \tau$ and $y=b \cos \tau$, where the parameter $\tau$ is measured clockwise from the positive $Y$-axis such that $\tau \equiv \text{ArcTan}\,\rho x/y$, not to be confused with the polar angle $\theta \equiv \text{ArcTan}\, x/y$; for a circle, $\rho=1$ (no eccentricity), so $\tau$ and $\theta$ coincide, while for $\rho\neq1$, $\tau=\theta$ only for $\theta= m \pi/2$. The length of an elementary elliptic arc is $ds=\sqrt{dx^2+dy^2}=\sqrt{a^2 \cos^2\!\tau + b^2 \sin^2\!\tau}d\tau$, therefore we introduce the non-linear curvilinear coordinate along the ellipse $s(\tau)=a \int_{0}^{\tau} \mathcal{I}(\tau')d\tau' =a \text{E}(\tau,e^2)$, where $\mathcal{I}(\tau)\equiv \sqrt{1-e^2\sin^2\!\tau} =\sqrt{\cos^2\!\tau+\rho^2\sin^2\!\tau}$ is the integrand of the \textit{incomplete} elliptic integral of the second kind, $\text{E}(\tau,e^2)$. $s(\tau)$ increases by one perimeter, $P=4aE(e^2)$, as $\tau$ changes by $2\pi$; for a thin circular ring with mean radius $r$, it would be $s(\tau)=s(\theta)=r \theta$. $E(e^2)\equiv E(\pi/2,e^2)$ is the {\it complete} elliptic integrals of the second kind of real argument $e^2\leq1$; typically, $E(e^2)$ is reported with the argument in the $[0,1]$ interval; however, when $\rho>1$,  $\text{E}(1\!-\!\rho^2)= \rho^{-1} \text{E}(1\!-\!\rho^{-2})$.

%
%

\section{The sine-Gordon modeling} 

Following Refs.\cite{gronbech,goldobin01}, a one-dimensional planar \textit{curved} JTJ of constant width in the presence of a barrier-parallel external magnetic field, ${\bf {H}}$, is described by the following partial differential equation for the Josephson phase $\phi$:

\vskip -5pt
\begin{equation}
\label{psge}
\phi_{\hat{s} \hat{s}} - \phi_{\hat{t} \hat{t}}-\sin \phi =\gamma(\hat{s}) + \alpha \phi_{\hat{t}}- \beta \phi_{\hat{s}\hat{s}\hat{t}}+ \frac{\Delta}{J_c \lambda_J } \frac{d H_\nu}{d\hat{s}}.
\end{equation}

\noindent In Eq.(\ref{psge}) the subscripts on $\phi$ denote partial derivatives; $\hat{s}$ is the distance, $s$, normalized to $\lambda_{J}$ and $\hat{t}$ is the time, $t$, normalized to $\omega _{p}^{-1}$, where $\omega_p/2\pi$, is the oscillation frequency of small amplitude waves. It is well known \cite{barone} that the Josephson lenght $\lambda_J$ gives a measure of the distance over which significant spatial variations of the phase occur. Further, $\gamma(\hat{s})$ is the local normalized bias current density.  The second and third terms in the right-hand side take into account the losses given by the normalized quasiparticle shunt loss coefficient $\alpha $ and the surface losses in the superconducting electrodes described by the surface loss coefficient $\beta$ \cite{barone}. $H_\nu$ is the component of the applied magnetic field normal to the \jun perimeter and $\Delta$ is the coupling between the external field and the field in the junction \cite{gronbech,PRB96}; it merely depends on the geometry and thickness of the junction electrodes. $J_c$ is the uniform maximum Josephson current density. Eq.(\ref{psge}) is called the Perturbed sine-Gordon Equation (PSGE). Because of its local form, it is quite general and holds for long one-dimensional \juns of any geometrical shape. It states that the magnetic field enters directly into the PSGE, in contrast to the case of linear junctions for which it appears only in the boundary conditions. For small fields a long linear JTJ behaves as a perfect diamagnet by establishing circulating screening currents which maintain the interior field at zero. This ''Meissner'' effect does not occur in long curved \juns where even a small magnetic field does penetrate the barrier; nevertheless, still the critical current decreases linearly with the applied field \cite{goldobin01,SUST15a}.

\vskip 5pt

\noindent In the absence of the right-hand side, the solitonic solution of Eq.(\ref{psge}) is a Josephson vortex (sine-Gordon kink), $\phi(\hat{s})= 4 \arctan \exp\left[(\hat{s}-\hat{s}_0)/ \sqrt{1-\hat{u}^2}\right]$, centered at $\hat{s}_0(\hat{t})$ and moving with constant normalized velocity \cite{scott} $\hat{u} = d \hat{s}_0/d\hat{t}$; $\hat{u}$ is the fluxon speed divided by the Swihart velocity, $c_0=\omega _{p}\lambda _J$. The right-hand side of Eq.(\ref{psge}) is usually considered as a perturbation \cite{scott}; it does not drastically change the vortex shape, but defines its dynamics, e.g., its equilibrium velocity. Such an approximation essentially treats the vortex as a rigid object, and its dynamics can be reduced to that of a relativistic underdamped point-like particle \cite{carapella,goldobin12}. As soon as $H_{\nu}\neq0$, the fluxon experiences a magnetic potential due to the interaction of its magnetic moment with the external magnetic field. If the normal field, $H_\nu$, changes slowly in comparison with the fluxon size, the normalized potential is $U_h({\hat{s}})=-\Delta h_{\nu}(\hat{s})$, where $h_{\nu}\equiv H_{\nu}/J_c \lambda_J$. In the well-known case of the ring-shaped junction, $h_\nu(s)\propto \cos(s/r)$, therefore the potential is sinusoidal; for an elliptic annulus we expect the potential to be still periodic and symmetric, but non-sinusoidal. The potential gradient, $F_h(\hat{s})\equiv -dU/d \hat{s}=\Delta dh_{\nu}/d\hat{s}$, gives the magnetic force exerted on the fluxon \cite{scott}; it is due to the interaction of the vortex magnetic moment with the external magnetic field and corresponds to the last term on the right-hand side of Eq.(\ref{psge}). In the small field limit, both $U_h$ and $F_h$ are proportional to the magnitude of the external field.

\subsection{PSGE for an elliptic annulus} 

The PSGE for an EAJTJ in the presence of a spatially homogeneous magnetic field, $H$, applied in the junction plane perpendicular to the $a$-semi-axis, has recently been derived \cite{SUST15a} with the assumption that the annulus width is constant and small compared to the \Jos penetration length:


$$\left(\frac{\lambda_J}{a\mathcal{I}}\right)^2 \left[ \phi_{\tau\tau} +  \frac{1-\rho^2} {\mathcal{I}^2} \sin\tau \cos\tau\, \phi_\tau \right]-\phi_{\hat{t}\hat{t}}-\sin \phi= $$
\vskip -5pt
\begin{equation}
= \gamma(\tau)  + \alpha \phi_{\hat{t}} - \left(\frac{\lambda_J}{a\mathcal{I}}\right)^2 \beta \left[ \phi_{\tau\tau \hat{t}} + \frac{1-\rho^2}{\mathcal{I}^2} \sin\tau \cos\tau\, \phi_{\tau \hat{t}}\right] +  h\Delta  \frac{\rho^2}{\mathcal{I}^4} \sin \tau,
\label{diff1}
\end{equation}

%

\noindent where $h=H/J_c\, a$; Eq.(\ref{diff1}) is supplemented by the periodic boundary conditions \cite{PRB96}:

\vskip -10pt
\begin{subequations}
\begin{eqnarray} \label{peri1}
\phi(\tau+2\pi)=\phi(\tau)+ 2\pi n,\\
\phi_\tau(\tau+2\pi)=\phi_\tau(\tau),
\label{peri2}
\end{eqnarray}
\end{subequations}

\noindent where $n$ is an integer number, called the winding number, corresponding to the algebraic sum of \Jos vortices (or fluxons) trapped in the \jun due to flux quantization in one of the superconducting electrodes. For $\rho=\mathcal{I}^2 (\rho,\tau)=1$, Eq.(\ref{diff1}) reduces to the well studied PSGE for ring-shaped junctions \cite{gronbech,PRB96}, while, in the limit $\rho \to 0$, it reproduces the classical PDE for a linear junction. Eq.(\ref{diff1}) states that the different sections of the annulus {\it feel} different fields; diametrically opposed points {\it feel} opposite fields and the field term in Eq.(\ref{diff1}) is out of phase with respect to the actual normal field. What matters in long EAJTJs is not the normalized perimeter, $\ell = P/\lambda_J= 4aE(e^2)/\lambda_J$, but rather the ratio $a/\lambda_J=\ell/4E(e^2)$. Once $a$ and $\lambda_j$ are given, a change in the annulus aspect ratio, $\rho$, corresponds to a variation in the length of the $b$ semi-axes. The commercial finite element simulation package COMSOL MULTIPHYSICS (www.comsol.com) was used to numerically solve Eq.(\ref{diff1}) subjected to cyclic boundary conditions Eqs.(\ref{peri1}) and (\ref{peri2}). In all calculations we set the damping coefficients $\alpha=0.1$ (weakly underdamped limit) and $\beta=0$, while keeping the current distribution uniform, $\gamma(\tau)=\gamma$. Strictly speaking, a uniformly distributed bias current would yield $\gamma(\tau)=\pi\gamma\mathcal{I}/2E(e^2)$; however, for small eccentricity ($e^2\simeq0$), it is $\pi\mathcal{I}(\tau,e^2)/2E(e^2)\approx1$. The uniformity of the current distribution in real samples is not a matter of concern in this work which does not take into account the geometry of the current carrying \jun electrodes.

\section{Fluxon dynamics} 

In this section we will consider the dynamic properties of a single Josephson vortex trapped in a current-biased EAJTJ. We first consider the case when no magnetic field is applied; later on we study the consequences of a periodic non-sinusoidal potential while varying the annulus eccentricity.

\subsection{The profile of the first Zero Field Step}

\noindent When cooling an annular \jun below its critical temperature one or more Josephson vortices (fluxons) may be spontaneously trapped in the \jun on a statistical basis. The corresponding flux quanta must be trapped in the superconducting loop formed by either the bottom or top electrode. The trapping probability is known to increase with the speed of the normal-to-superconducting transition \cite{PRB06,PRB08}. After a successful trapping procedure the \jun zero-voltage critical current is considerably smaller and a stable finite-voltage current branch, called zero-field step (ZFS), appears in the junction current-voltage characteristic indicating that the bias current forces the fluxon(s) to travel along the annulus in the absence of collisions. In the simplest picture the velocity of a single fluxon, considered as a relativistic particle, is determined by a balance between the driving force on the fluxon and the drag force due to the dissipative losses. In the absence of a magnetic field, the driving force is proportional to the bias current density which is assumed uniform in the junction. A relativistic formula for the fluxon motion was derived \cite{scott} for an infinitely long junction using a perturbative approach. It predicts a smooth current-voltage profile of the ZFS branch in annular junctions where, as on an infinite line, the fluxon moves only subjected to cyclic boundary conditions: 


\begin{equation}
\gamma(\hat{u})=\frac{4\alpha/\pi}{\sqrt{\hat{u}^{-2}-1}}.
\label{gammau}
\end{equation}

\begin{figure}[tb]
\centering
\includegraphics[width=7cm]{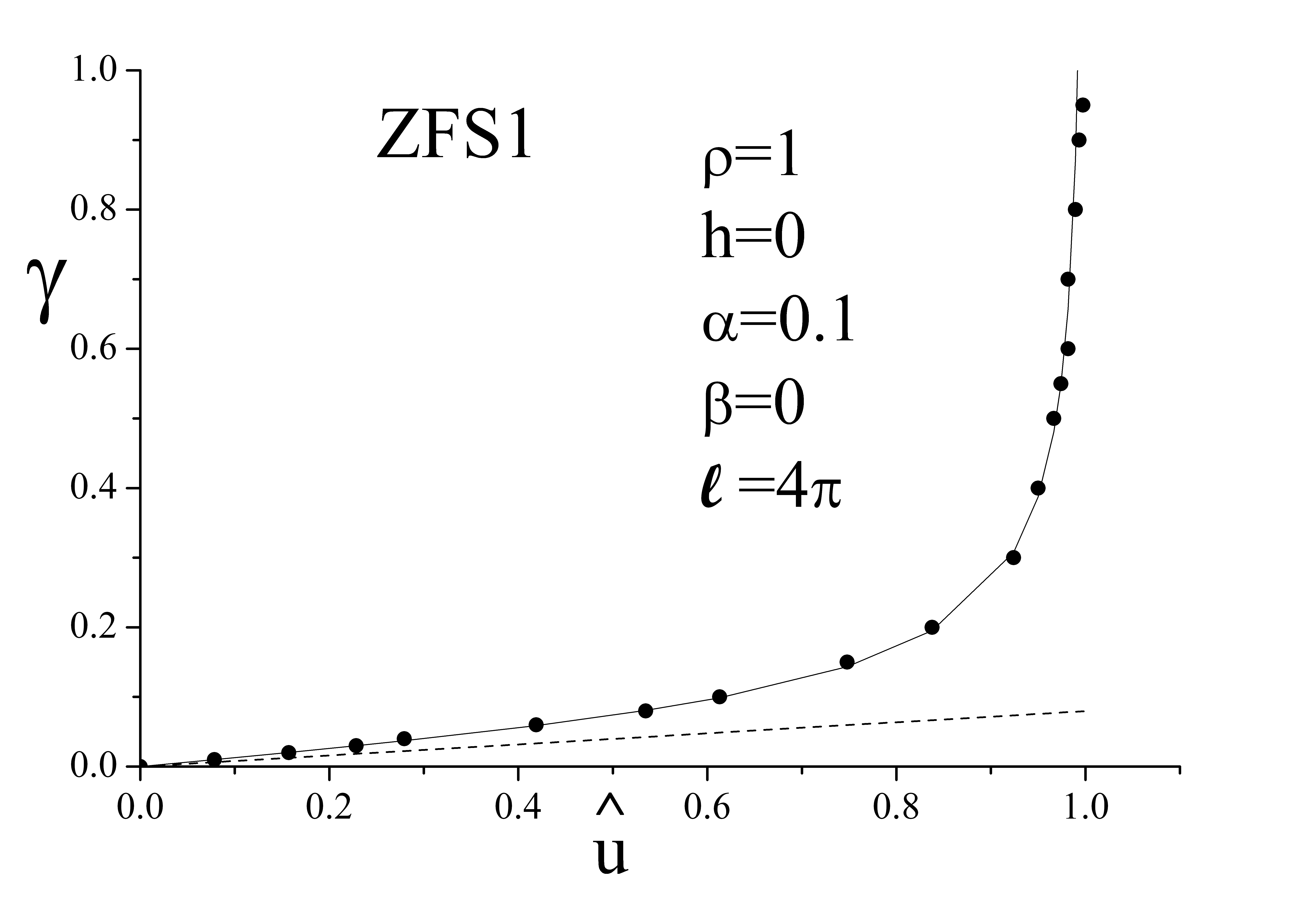}
\caption{The dots refer to the numerically computed profile of the first zero-field step for an EAJTJ with no external magnetic field; the profile is $\rho$-independent. Results are calculated integrating Eq.(\ref{diff1}) with $l=4\pi$, $\rho=1$, $\alpha=0.1$, $\beta=0$, $h=0$, and $n=1$. The solid line is the perturbative model expectation in Eq.(\ref{gammau}). The dashed line depicts the quasi-particle current $\gamma/\ell \alpha$.}
\label{ZFS1h0}
\end{figure}

\noindent The dots in Figure~\ref{ZFS1h0} show the numerically computed current-voltage (i.e., $\gamma$ versus $<\phi_{\hat{t}}>$) characteristic of a ring-shaped \jun of normalized length $\ell=4\pi$ with one trapped fluxon ($n=1$). Noticing that the mean voltage generated by a fluxon moving with velocity $\hat{u}$ is given by $V \propto <\phi_{\hat{t}}>=2\pi \hat{u}/\ell$, and since, from Eq.(\ref{psge}), $\gamma$ means a force, we can think of the plot in Figure~\ref{ZFS1h0} also as force-velocity characteristics. The normalized velocity, $\hat{u}$, was determined from the fluxon revolution period $\hat{T}$, as $\hat{u}=\ell/\hat{T}$ and was defined to be positive for fluxons rotating clockwise. The dashed line depicts the quasi-particle current $\gamma/\ell \alpha$. The profile of the ZFS smoothly goes to zero as the bias current goes to zero and is point-symmetric for negative biases (not shown), meaning that the fluxon rotates counterclockwise. The ZFS profile is insensitive to the annulus eccentricity as far as we keep its perimeter constant; more specifically, numerical simulations showed that the fluxon travels at the same stationary power-balance speed regardless of the annulus aspect ratio, i.e., the revolution period only depends on the bias current. The solid line in Figure~\ref{ZFS1h0} is the $\gamma(\hat{u})$ dependence according to Eq.(\ref{gammau}); the good agreement indicates that the perturbative approach also apply to a fluxon traveling along an elliptic annulus.


\subsection{Fluxon depinning currents}

\begin{figure}[tb]
\centering
\subfigure[ ]{\includegraphics[width=7cm]{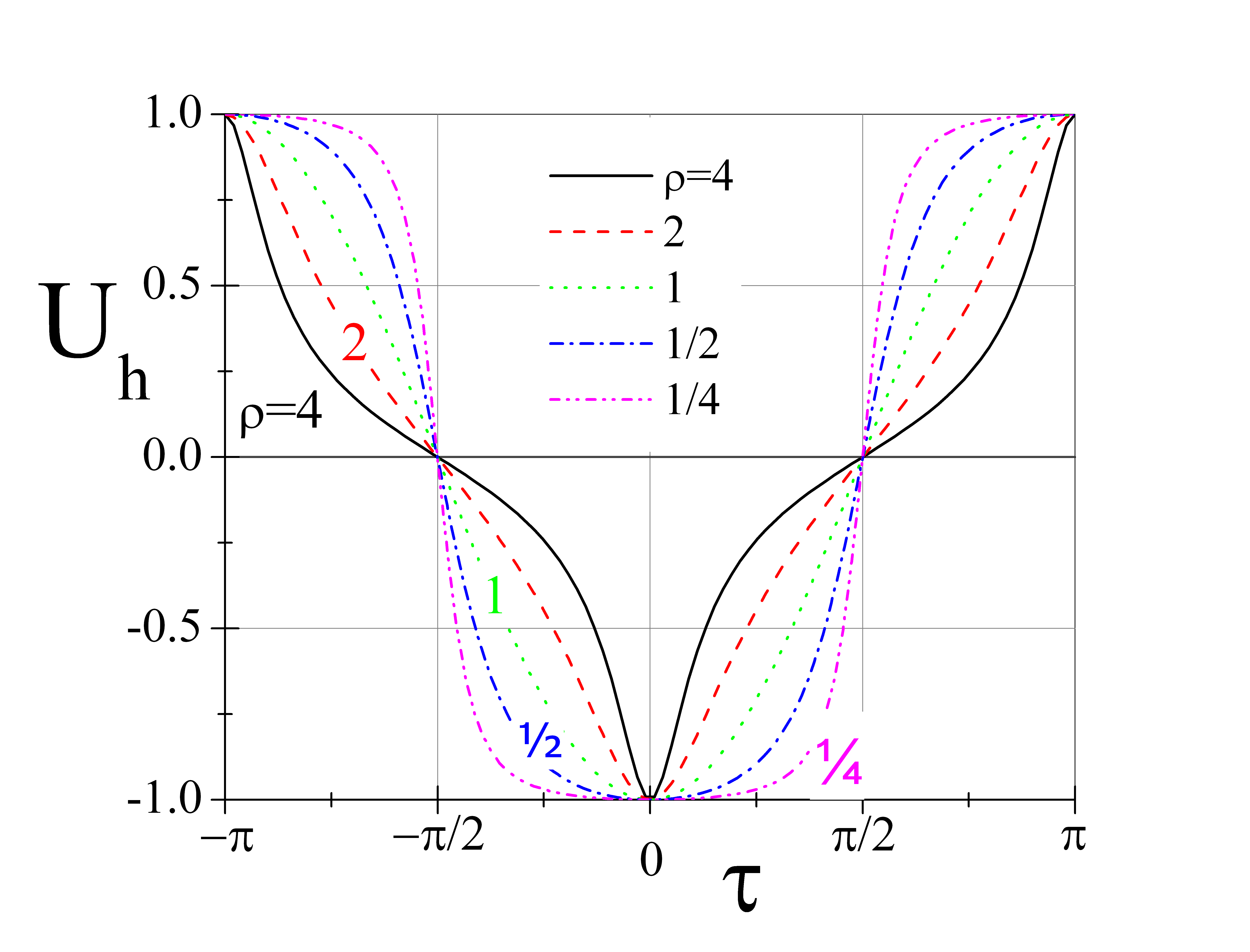}}
\subfigure[ ]{\includegraphics[width=7cm]{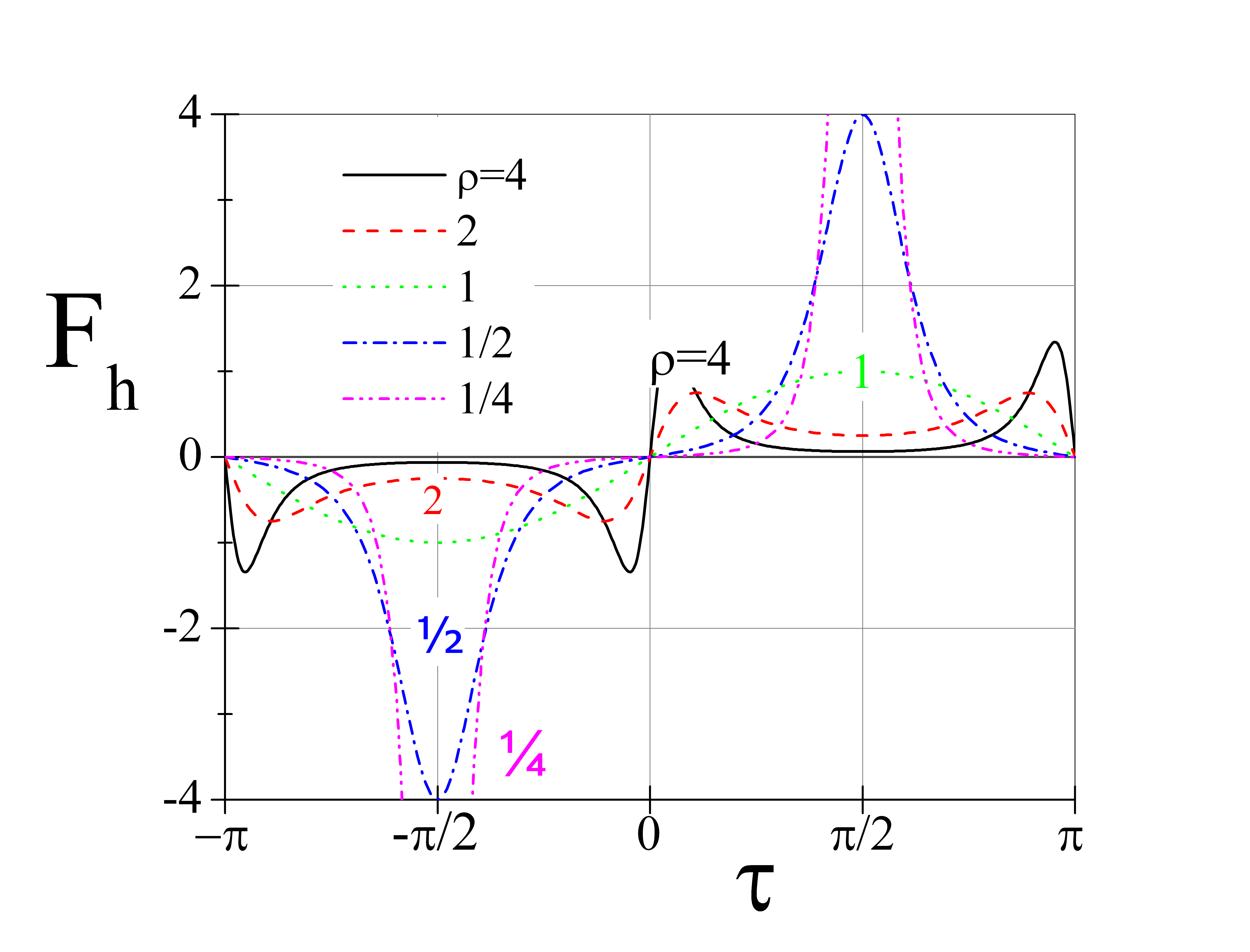}}
\caption{(Color online)(a) Plots of the magnetic potential $U_h=\mathcal{I}^{-1}(\tau) \cos\tau$ for several $\rho$ values ($\Delta=h=1$); (b) as in (a) for the forcing term $F_h=\rho^2 \,\mathcal{I}^{-4}(\rho,\tau) \sin\tau$.}
\label{forcing}
\end{figure}

In the presence of a magnetic field the fluxon is trapped in a potential well until the Lorentz force associated with the bias current is strong enough to start its motion. The potential shape drastically depends on the annulus eccntricity and so does the depinning current, $\gamma_d$. Considering that ${\bf {H}}=(0,H)$, the normal field along the annulus perimeter is \cite{SUST15a} $H_\nu(\tau)= H \mathcal{I}^{-1}(\tau) \cos\tau$. By assuming that the annulus is long enough so that the left and right tails of the fluxon do not interact, it turns out that the magnetic potential for an EAJTJ is $U_h(\tau)=-\Delta h_{\nu}(\tau)=-\Delta h \mathcal{I}^{-1}(\tau) \cos\tau$ and is plotted in Figure~\ref{forcing}(a) for several $\rho$ values (for $\Delta=h=1$). It is seen that the potential is periodic and symmetric and vanishes at the equatorial points, $\tau=\pm \pi/2$ (it is sinusoidal only for $\rho=1$). With our settings, the potential well occurs at the origin, $s=\tau=0$.  In the limit $\rho \to 0$, $U_h(\tau)$ approaches a symmetric square wave profile, while it resembles a train of alternate unitary pulses, in the opposite limit, $\rho>>1$. The directional derivative of the normal field is ${d h_{\nu}}/{d\hat{s}} =(\lambda_j/a) h_\parallel  \rho^2 \mathcal{I}^{-4}  \sin\tau$, therefore, the potential force experienced by a fluxon is a function of the vortex location: $F_h\equiv -dU_h/d\hat{s}=\Delta dh_{\nu}/d\hat{s}= (\lambda_j/a)\Delta h_\parallel \rho^2 \mathcal{I}^{-4}  \sin\tau$; $F_h(\tau)$ is plotted in Figure~\ref{forcing}(b) for different $\rho$ values (for $\lambda_j/a=\Delta=h=1$). We found that as far as $\rho\leq 2/\sqrt{3}\simeq 1.154$, $F_h(\tau,\rho)$ is maximum for $\tau=\pi/2$; increasing $\rho$ a twofold maximum develops at $\bar{\tau}=\cos^{-1}\pm\sqrt{(3 \rho^2-4)/(3 \rho^2-3)}$ that approaches the values $0$ and $\pi$ for very large aspect ratios. The fluxon depinning occurs when the bias current, $\gamma$, just exceeds the limit of the depinning force: 

\vskip -5pt
\begin{equation}
\label{cases}
F_{d}(\rho)=\begin{cases}{F_h \left(\frac{\pi}{2},\rho \right)= h\Delta \rho^{-2}} &\mbox{if } 0\leq\rho\leq \frac{2}{\sqrt{3}} \\
{F_h \left(\cos^{-1} \sqrt{\frac{3 \rho ^2-4}{3 \rho ^2-3}},\rho\right)=h\Delta \frac{3^{3/2}\,\,\rho^2}{2^4\sqrt{\rho^2-1}}  } & \mbox{if } \rho\geq \frac{2}{\sqrt{3}}.
\end{cases}
\end{equation}

\begin{figure}[tb]
\centering
\includegraphics[width=8cm]{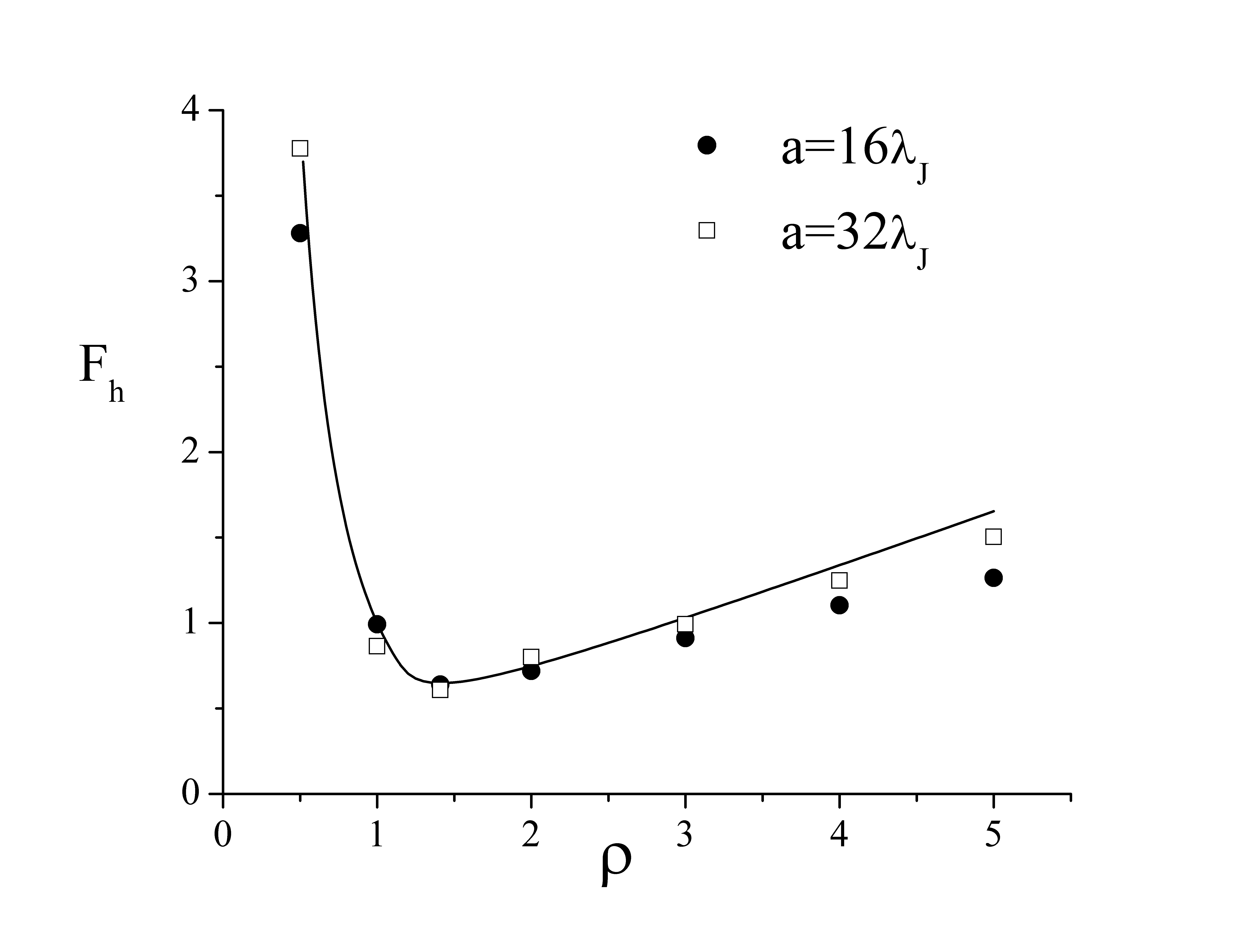}
\caption{Depinning force $F_d$ versus the annulus axes-ratio $\rho$ for $\Delta=h=1$; the solid lines refers to Eq.(\ref{cases}), while the solid circles and the open squares are the numerically computed values for elliptic annuli having the semi-axes $a=16\lambda_j$ and $a=32\lambda_j$, respectively.}
\label{depinning}
\end{figure}

\noindent $F_{d}(\rho)/ h\Delta$ is shown by the line in Figure~\ref{depinning}; it has a minimum when $\rho=\sqrt{2}$. For small aspect ratios it grows as $\rho^{-2}$, while it increases linearly in the opposite limit. The solid circles and the open square are the numerically computed values of $\gamma_d$ for, respectively, $a=16 \lambda_J$ and $a=32 \lambda_J$, setting $h\Delta=\lambda_J/a$ in Eq.(\ref{diff1}). We see that the longer the annulus perimeter is, the better is the agreement with the infinite length expectation. Further, the linearity of the depinning current for small field values was also successfully tested. 

\noindent Once the fluxon has been unpinned its speed is periodically modulated by the magnetic potential. However, due to the symmetry of the potential and the ohmic losses included in the sine-Gordon modeling, its average value does not changes. The more realistic viscous losses proportional to $\phi_t^2$ rather than $\phi_t$ would yield a different results \cite{PRB98}.

\begin{figure}[tb]
\centering
\includegraphics[width=8cm]{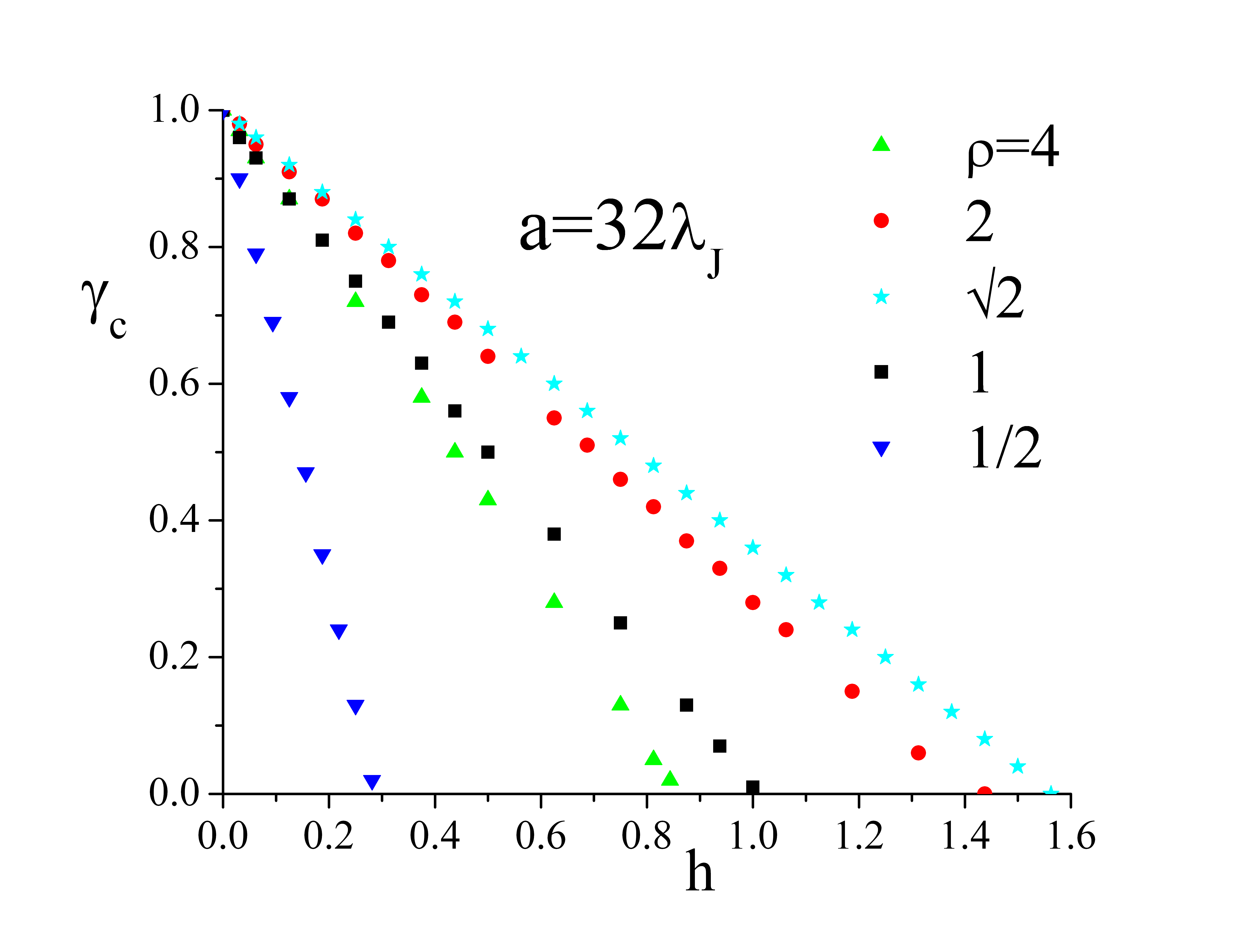}
\caption{(Color online) Calculated magnetic diffraction patterns of long EAJTJs for several values of the axis ratio, $\rho$ when $a=32\lambda_j$ and $\Delta=1$. The magnetic field is normalized to $J_c a$.}
\label{MDP}
\end{figure}

\noindent $\sqrt{2}$ is a peculiar value of $\rho$ for which the effect of a given applied field on an EAJTJ is minimum. This can be seen  also in the absence of trapped flux, by considering how the external field modulates the critical current, $\gamma_c$, above which the junction switches to a a finite voltage state. Figure~\ref{MDP} shows the main lobe of the magnetic diffraction patterns for several $\rho$ values when $a=32\lambda_J$. We see that the largest critical field is achieved when $\rho=\sqrt{2}$.

\section{Egg-shaped annular junctions}

\begin{figure}[tb]
\centering
\subfigure[ ]{\includegraphics[width=7cm]{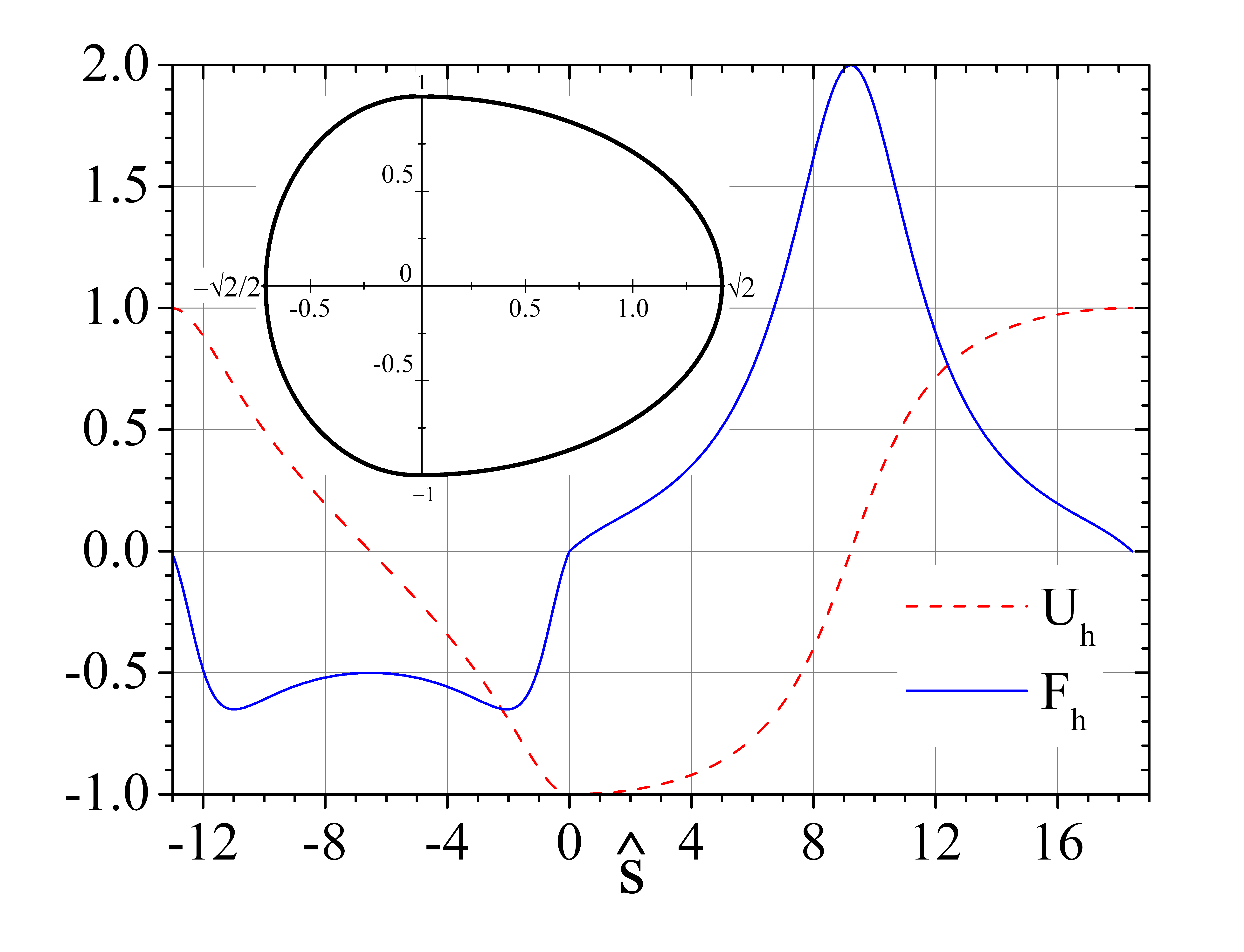}}
\subfigure[ ]{\includegraphics[width=7cm]{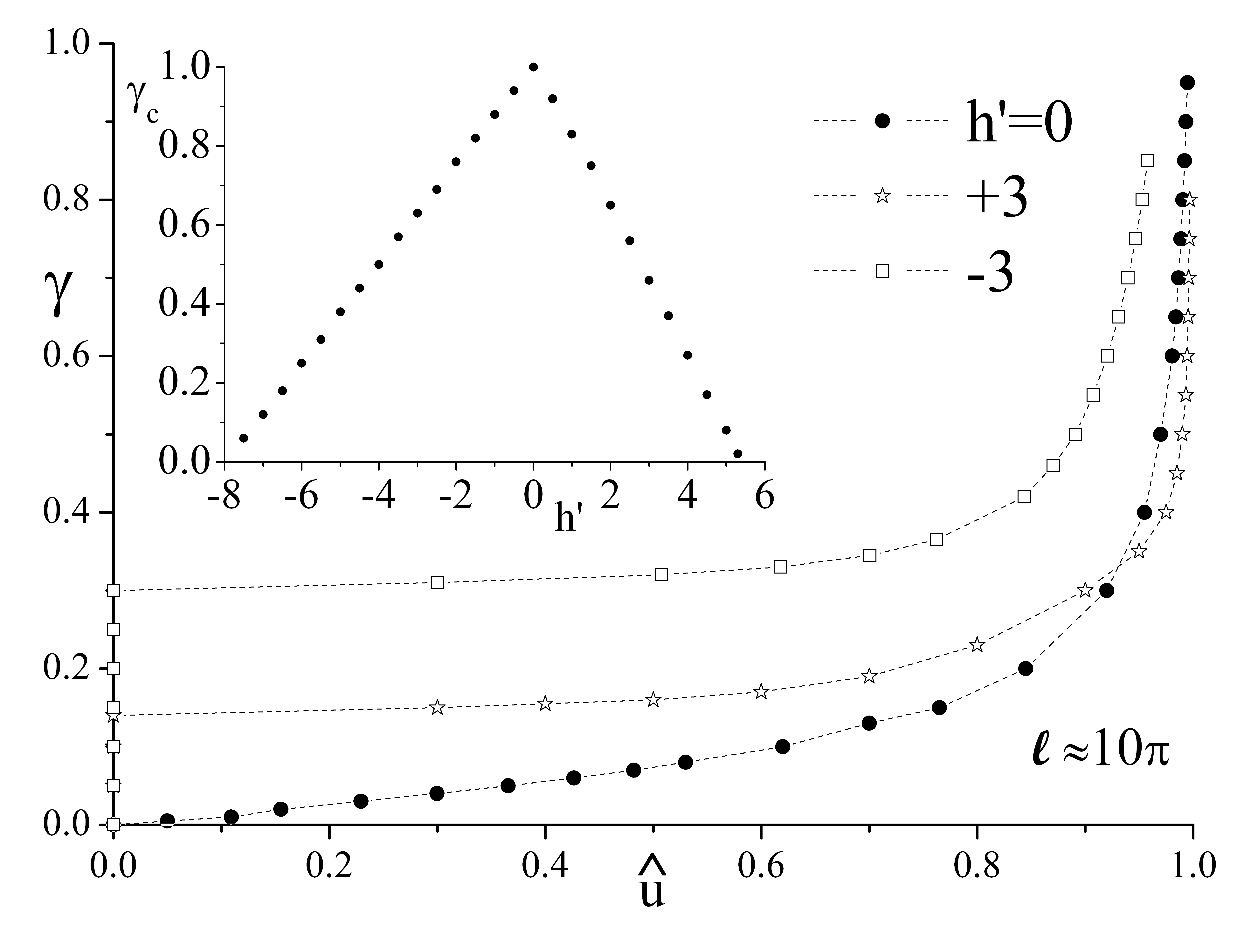}}
\caption{(Color online)(a) The magnetic potential $U_h$ and forcing $F_h$ for the egg-shaped annular \jun shown in the inset; $\hat{s}=s/\lambda_j$ is the normalized distance measured from the top. (b) Numerically computed ZFS profiles for three different values of the normalized magnetic field, $h'=H/J_c \lambda_J$. The inset shows the main lobe of the computed magnetic diffraction pattern.}
\label{doubleplot}
\end{figure}

\noindent Asymmetric field profiles in long JTJs were recently investigated \cite{carapella,goldobin01,wallraf3} in order to exploit the rectifying property \cite{magnasco} of a ratchet potential. Playing with the shape of the tunnel barrier it is possible to generate asymmetric periodic potentials  by simply applying an uniform magnetic field. It is easy to recognize that the asymmetric tapered oval configuration depicted in the inset of Figure~\ref{doubleplot}(a) implements a deterministic ratchet potential. It consists of two semi ellipses having the same $b$ semi-axis and different $a$ semi-axis; to the left is $a_l=\sqrt{2}b/2$ and to the right $a_r=2a_l=\sqrt{2}b$. In this way the oval assembles the cross section of an hen egg. At variance with the other geometries previously used to study the effect of non symmetric potentials \cite{goldobin01, wallraf3}, the fluxon dynamics in an asymmetric elliptic oval is described by an analytic PSGE; in fact, Eq.(\ref{diff1}) has been numerically solved	 by setting $a(\tau)=a_l$ and $\rho(\tau)=\sqrt{2}/2$ for $\tau\in[-\pi,0]$ and  $a(\tau)=a_r$ and $\rho(\tau)=\sqrt{2}$ for $\tau\in[0,\pi]$. We set $b=4.83\lambda_J$ so that the total annulus perimeter is $L=2 a_lE(1/2)+2a_r E(-1)=\sqrt{2}bE(1/2)(1+\sqrt{2})\approx 10 \pi \lambda_j$. Since the magnetic potential is averaged over the vortex size ($\propto \lambda_j$), the smooth changes in the oval curvature allows us to apply perturbation theory. The magnetic potential and the forcing term generated by an uniform in-plane magnetic field applied perpendicular to the symmetry axis are shown in Figure~\ref{doubleplot}(a), respectively, by the dashed and solid line. Eq.(\ref{tau}) has been used to express $U_h$ and $F_h$ as a function of the normalized distance $\hat{s}=s/\lambda_j$. We see that on the average the accelerating force in the right side is about twice larger than the mean decelerating one in the left; in addition, the positive force acts over a longer distance. 

\subsection{Deterministic ratchet potential}

\noindent Figure~\ref{doubleplot}(b) shows the computed step profile for three values of the external magnetic field $h'=H/J_c \lambda_J$, namely $3$, $0$ and $-3$. According to Eq.(\ref{diff1}) a reversal of the magnetic field is equivalent to a current reversal and vice versa. The full dots show the ZFS profile of the egg-shaped annular JTJ with one trapped fluxon when the ratchet potential is off ($H=0$); again the fit with the perturbation result of Eq.(\ref{gammau}) was excellent. When a positive ratchet field is turned on (open stars) we see that the current amplitude of the step is reduced and a current range with zero-mean voltage (zero velocity) appears. This hysteretic behavior reflects the relevance of the inertial effects typical of underdamped systems. The zero-voltage current is the fluxon trapping current, i.e., the minimum current at which a fluxon still moves along the system, not being trapped by the potential. The situation drastically changes when the magnetic field in reverted (open squares); as expected, the average fluxon speed is much smaller and the hysteresis more pronounced.

\noindent The asymmetry of the potential has been also probed by calculating the main lobe of the magnetic diffraction pattern for the egg-shaped geometry; as shown in the inset of Figure~\ref{doubleplot}(b), the slopes for positive and negative fields are quite different.

\section{Conclusions}

In this paper we investigated the dynamic properties of a not simply connected planar \Jos tunnel \jun shaped as an elliptic annulus of constant width (which does not imply that the barrier area is delimited by two confocal ellipses). This system can be analyzed in the contest of the theory developed for curved one-dimensional long \Jos tunnel \juns \cite{goldobin01} provided cyclic boundary conditions are imposed; the perturbed sine-Gordon equation for the \Jos phase appropriate to \elli \ann \Jos tunnel \juns was derived in a recent paper \cite{SUST15a}. Extensive numerical simulations showed that, as far as no magnetic field is applied, the single fluxon dynamics in a current biased \elli annulus is not affected by its eccentricity; the stationary fluxon speed only depends on the balance between the Lorentz force induced by the bias current and the dissipation drag acting on the vortex.

\noindent Elliptic annular junctions inherently have specular symmetry with respect to their principal axes: quite obviously an in-plane magnetic field breaks the system symmetry along its direction. If the field is applied along one semi-axis the induced periodic potential is symmetric and non-sinusoidal. We studied the trapping properties of the potential as a function of the ellipse aspect ratio and found that smallest effect is achieved when $\rho=\sqrt{2}$. We also suggested an egg-shaped geometry with only one axis of symmetry; here the magnetic field coupled to the annulus lacks reflection symmetry, so accomplishing a rectifying potential in which a soliton is preferentially accelerated in one half of the junction perimeter and accelerated in the opposite part. In passing, we note that an asymmetric potential can simply be implemented by applying the uniform in-plane field at a generic angle $\theta\neq k\pi/2$ measured from the $b$ semi-axis of an EAJTJ, ${\bf {H}}=(H\sin \theta,H\cos\theta)$; in this case the normal field along the annulus perimeter would be $H_\nu(\theta,\tau)/ H =(a \cos\theta\cos\tau + b\sin\theta\sin\tau) / \sqrt{a^2\cos^2 \tau +b^2\sin^2 \tau}$. However, in such a case the ratchet effect would be much less pronounced than for the asymmetric oval case. 


\section*{Acknowledgments}
\noindent RM and JM acknowledge the support from the Danish Council for Strategic Research under the program EXMAD.

\renewcommand{\theequation}{A-\arabic{equation}}
\setcounter{equation}{0}  
\section*{APPENDIX $\tau(s)$} 

\noindent If one wants to express magnetic potential, $U_h$, or force, $F_h$, as a function of the more intuitive curvilinear coordinate, $s$, then the dependence $\tau(s)$ is needed. Unfortunately, the \elli integrals of the second kind are not invertible in terms of single-valued functions (at variance with the \elli integrals of the first kind). In Ref.\cite{AS} a series expansion of $\tau$ in powers of $\text{E}(\tau,m)$ is reported, but only in the limit of small $\text{E}$. Another efficient ansatz is a higher order Newton-Raphson root-finding method, since the derivatives of $\text{E}(\tau,e^2)$ with respect to $\tau$ are well known: $\partial \text{E}(\tau,e^2)/ \partial \tau = \sqrt{1-e^2\sin^2 \tau}$ and $\partial^2 \text{E}(\tau,e^2)/\partial \tau^2 = e^2\sin2\tau (2\partial E/\partial \tau)^{-1}$; for this one needs the implementation of an iterative algorithm and, in addition, a solid computation of the original $\text{E}(\tau,e^2)$ itself. 

\begin{figure}[bt]
\centering
\includegraphics[width=9cm]{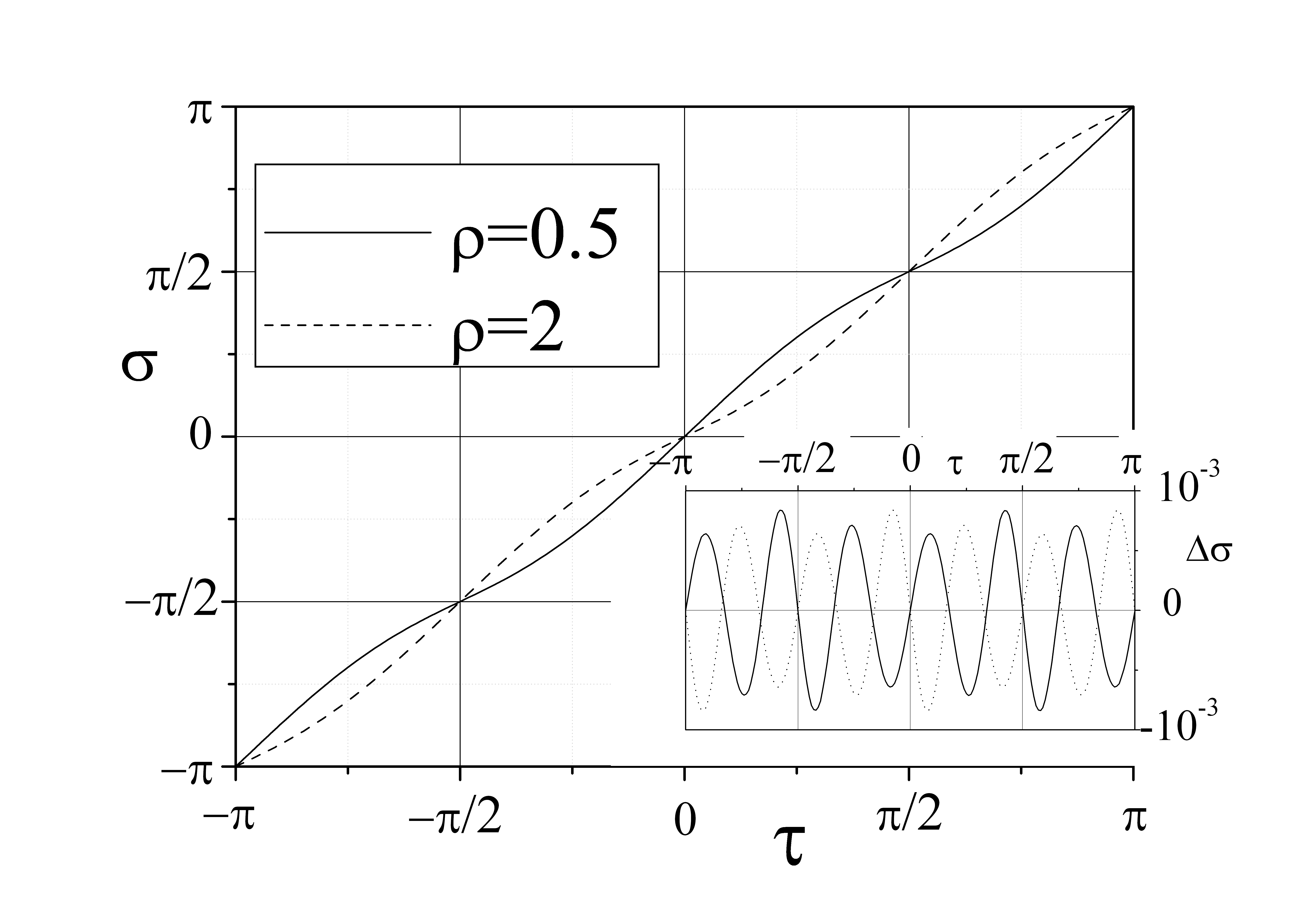}
\caption{Curvilinear coordinate normalized to the ellipse circumference $\sigma(\tau,\rho)=\pi \text{E}(\tau,1-\rho^2)/ 2\text{E}(1-\rho^2)$; in the inset $\Delta \sigma(\tau)=\sigma-\sigma'$ where $\sigma'(\tau,e^2)\equiv\tau+A_2(e^2) \sin2\tau + A_4(e^2)\sin4\tau$.}
\label{sigma}
\end{figure}

\noindent To circumvent these difficulties, let us consider, with no loss of generality, an ellipse of generic eccentricity, $e^2=1-\rho^2$, and length $2\pi$, whose curvilinear coordinate $\sigma(\tau,e^2)=2\pi s(\tau,e^2)/L=\pi \text{E}(\tau,e^2)/ 2\text{E}(e^2)$ simply scales with the distance $s$ measured on a ellipse having the same eccentricity and generic length $L$. Figure~\ref{sigma} shows the $\tau$-dependence of $\sigma(\tau,e^2)$ for two reciprocal values of the axis ratio $\rho=\sqrt{1-e^2}$, namely, $1/2$ and $2$. We see that the incomplete elliptic integrals of the second kind are quasi-periodic functions with respect to $\tau$, namely, $\text{E}(\tau+k\pi,e^2)= \text{E}(\tau,e^2)+ 2k\text{E}(e^2)$; in terms of the the normalized distance, $\sigma$, it is $\sigma(\tau+k\pi,e^2)= \sigma(\tau,e^2)+k\pi$. In other words, $\sigma(\tau,e^2)$ is a linear function in $\tau$ plus a $\pi$-periodic oscillation, and a Fourier expansion of the difference $\sigma(\tau,e^2)-\tau$ with one or two terms is sufficient when the ellipse is not extremely eccentric; accordingly we can write: $\sigma(\tau,e^2)\simeq \tau+A_2(e^2) \sin2\tau + A_4(e^2)\sin4\tau \equiv \sigma'(\tau,e^2)$, where, in general \cite{note3}, $A_2(1/\rho)=-A_2(\rho)$ and $A_4(1/\rho)=A_4(\rho)$, while, for $0\leq\rho\leq1$,	 $A_2(\rho)\approx (1-\rho)/3$ and $A_4(\rho)\approx (1-\sqrt{\rho})/30$. The inset of Fig.~\ref{sigma} shows the difference $\Delta \sigma=\sigma-\sigma'$ whose absolute value is always less than $10^{-3}$. We also note that, to a very high degree of approximation,  $\tau(\sigma,\rho) \simeq \sigma(\tau,1/\rho)$, in other words, $\sigma(\tau)$ can be inverted by simply operating the transformation $\rho \to 1/\rho$, namely,

\begin{equation}
\tau(\sigma,\rho) \simeq \sigma + A_2(1/\rho) \sin2\sigma + A_4(1/\rho)\sin4\sigma=\sigma - A_2(\rho) \sin2\sigma + A_4(\rho)\sin4\sigma.
\label{tau}
\end{equation}

\noindent This empirical finding provides an overall accuracy better than $0.5$\% for $\rho=0.5$ and $2$; the accuracy improves a lot for intermediate values. For an \elli annulus of physical length $L$, one has to remind that $\sigma=2\pi s /L$. 
%


\newpage
\begin{thebibliography}{99}

\bibitem{scott} D. W. McLaughlin and A. C. Scott, {\it Phys. Rev. A} {\bf 18}, 1652 (1978); A. V. Ustinov, {\it Physica D}{\bf 123}, 315 (1998).
  
\bibitem{davidson}  A. Davidson, B. Dueholm, B. Kryger, and N. F. Pedersen,
Phys. Rev. Lett. {\bf 55}, 2059 (1985).

\bibitem{dueholm}  A. Davidson, B. Dueholm, and N. F. Pedersen, J. Appl. Phys. 
{\bf 60,} 1447 (1986).

\bibitem{hue} A. V. Ustinov, T. Doderer, R. P. Huebener, N. F. Pedersen, B. Mayer, and V. A. Oboznov, Phys. Rev. Lett. {\bf 69}, 1815 (1992).

\bibitem{gronbech} N. Gr\"{o}nbech-Jensen, P. S. Lomdahl, and M. R. Samuelsen, {\it Phys. Lett. A} {\bf 154}, 14 (1991); N. Gr\"{o}nbech-Jensen, P. S. Lomdahl, and M. R. Samuelsen, {\it Phys. Rev. B} {\bf 43}, 12799 (1991).

\bibitem{ustinov}  A.V. Ustinov, JETP Lett., {\bf 64 }191 (1996).

\bibitem{PRB98}  N. Martucciello, J. Mygind, V.P. Koshelets, A.V.
Shchukin, L.V. Filippenko and R. Monaco, {\it Phys. Rev. B} {\bf 57}, 5444
(1998).

\bibitem{wallraf3} A. Wallraff, A. Lukashenko, J. Lisenfeld, A. Kemp, M.V. Fistul, Y. Koval and A. V. Ustinov, {\it Nature} {\bf 425}, 155 (2003).

\bibitem{magnasco} M.O. Magnasco, {\it Phys. Rev. Lett.} {\bf 71}, 1477 (1993).

\bibitem{SUST15a} R. Monaco, C. Granata, A. Vettoliere, and J. Mygind, {\it Supercond. Sci. Technol.} {\bf 28}, 085010 (2015).

\bibitem{http} http://mathworld.wolfram.com/EllipseParallelCurves.html

\bibitem{goldobin01} E. Goldobin, A. Sterck, and D. Koelle, {\it Phys. Rev. E} {\bf 63}, 031111 (2001).

\bibitem{barone}  A. Barone and G. Patern\`o, {\em Physics and Applications of the Josephson Effect }(Wiley, New York, 1982).

\bibitem{PRB96} N. Martucciello, and R. Monaco, {\it Phys. Rev. B} {\bf 53} 3471 (1996); N. Martucciello, C. Soriano and R. Monaco, {\it Phys. Rev. B} {\bf 55} 15157 (1997).

\bibitem{carapella} G. Carapella, {\it Phys. Rev. B} {\bf 63}, 054515 (2001); G. Carapella, and G. Costabile {\it Phys. Rev. Lett.} {\bf 87}, 077002 (2001). 

\bibitem{goldobin12} M. Knufinke, K. Ilin,M. Siegel, D. Koelle, R. Kleiner, and E. Goldobin, {\it Phys. Rev. E} {\bf 85}, 011122 (2012).

\bibitem{PRB06}  R. Monaco, M. Aaroe, J. Mygind, R.J. Rivers, and V.P. Koshelets,  {\it Phys. Rev. B} {\bf 74}, 144513 (2006).

\bibitem{PRB08}  R. Monaco, M. Aaroe, J. Mygind, R.J. Rivers, and V.P. Koshelets,  {\it Phys. Rev. B} {\bf 77}, 054509 (2008) and references therein.

\bibitem{AS} Abramowitz, Milton; Stegun, Irene A., eds. (1972), Handbook of Mathematical Functions with Formulas, Graphs, and Mathematical Tables, New York: Dover Publications, ISBN 978-0-486-61272-0. As outlined in chapter 3.6.25 the following series expansion holds: $\tau=\text{E}(\tau,m)+(m/6)\text{E}^3(\tau,m)+ (m/120)(13m-4)\text{E}^5(\tau,m)+(m/5040)(493m^2-284m+16)\text{E}^7(\tau,m)+(m/362880)(37369m^3-31224m^2+4944m-64)\text{E}^9(\tau,m) ...$; the expansion can be obtained by the Mathematica command In[1]:=InverseSeries[Series[EllipticE[$\tau$, m], {$\tau$, 0, 10}]].

\bibitem{note3} The analytic expressions of the Fourier coefficients are: 

$A_2(e^2)= \left[(2-e^2)- 2(1-e^2)\text{K}(e^2)/\text{E}(e^2)\right]/3e^2$ and 

$A_4(e^2)=\left[-\left(16-16 e^2+e^4\right)+ 8 \left(2-3 e^2+e^4\right)\text{K}(e^2)/ \text{E}(e^2)\right]/30e^4$, 

where $\text{K}(e^2)$ is the complete \elli integral of the first kind. 




























 

























%






























\end{thebibliography}
\end{document}